# HeTraX: Energy Efficient 3D Heterogeneous Manycore Architecture for Transformer Acceleration

Pratyush Dhingra, Janardhan Rao Doppa, and Partha Pratim Pande.
Washington State University, Pullman WA, USA. {pratyush.dhingra, jana.doppa, pande}@wsu.edu

## ABSTRACT

Transformers have revolutionized deep learning and generative modeling to enable unprecedented advancements in natural language processing tasks and beyond. However, designing hardware accelerators for executing transformer models is challenging due to the wide variety of computing kernels involved in the transformer architecture. Existing accelerators are either inadequate to accelerate end-to-end transformer models or suffer notable thermal limitations. In this paper, we propose the design of a three-dimensional heterogeneous architecture referred to as HeTraX specifically optimized to accelerate end-to-end transformer models. HeTraX employs hardware resources aligned with the computational kernels of transformers and optimizes both performance and energy. Experimental results show that HeTraX outperforms existing state-of-the-art by up to 5.6x in speedup and improves EDP by 14.5x while ensuring thermally feasibility.

## 1 INTRODUCTION

Transformers have emerged as a focal point in the field of deep learning [1]. However, the size of transformer models is continuously increasing, driven by the pursuit of higher accuracy and improved capabilities across various natural language processing (NLP) tasks [1]. This trend of ever-increasing model size has given rise to new challenges in terms of memory and compute requirements. Conventional computing platforms, such as GPUs, suffer from suboptimal performance due to the memory demands imposed by the models with millions/billions of parameters. Consequently, this has motivated the need to explore and develop energy-efficient accelerators for transformers.

The transformer architecture is primarily characterized by a self-attention mechanism and a feed-forward network [1]. Transformers include multiple self-attention heads (multi-head attention), with each head operating in parallel. Subsequently, the feed-forward (FF) network is employed, which includes multiplication with the trainable weights. The end-to-end transformer model also consists of additional computations such as softmax, layer-normalization, activation function, positional encoding, etc. These computational kernels give rise to the heterogeneity of operations in the transformer architecture. Recently, processing-in-memory (PIM) has emerged as a promising approach to accelerate the training/inference of deep neural networks (DNNs) [2]. Emerging resistive random-access memory (ReRAM)-based PIM architectures can achieve higher performance and better energy efficiency than GPU-based counterparts [2]. However, the attention mechanisms of transformers, with the need for dynamic operand multiplications, would necessitate a high frequency of write operations to ReRAM cells. It is well known that ReRAM writes are slow and suffer from limited write endurance [3]. Hence, ReRAMs are not suitable for multi-head attention (MHA) computation in transformers. Conversely, ReRAMs remain a viable candidate for executing the FF network within transformers, which perform multiplications with learned weights (stationary weights). Considering the diverse nature of computational kernels, it becomes necessary to consider appropriate heterogeneity in the computing resources to develop manycore hardware accelerators for transformers. Hence, we propose to use streaming multiprocessors (SMs) along with the associated memory controllers (MCs) for the computation of MHA and ReRAM-based PIM for the FF network. This approach represents an allocation of hardware resources well-aligned with the computational kernels of transformers. It is a departure from the current practice where most of the existing transformer accelerators utilize homogenous hardware platforms (ReRAM-based PIM, DRAM, GPU, etc.)

Three-dimensional (3D) integration is a viable solution for heterogeneous computing since it can support various types of cores fabricated using different technologies in a single system. Additionally, vertical links help to enhance the overall system performance and concurrently reduce inter-core data transfer costs. However, it is essential to recognize that 3D architecture introduces significant challenges in terms of thermal management. Thus, a thermal-aware approach becomes imperative during the design of any such architecture. Additionally, the complex data flow in the transformer model makes the underlying framework of integrating the heterogeneous cores non-trivial, necessitating careful optimization to facilitate efficient data exchange. In this work, we propose a 3D-heterogenous architecture designed for low-latency and energy-efficient transformer inference. Specific contributions of this work are :

1. We propose a 3D-heterogenous architecture referred to as HeTraX that incorporates distinct planar tiers where each tier is tailor-made for either MHA or the FF network.
2. HeTraX synergistically leverages the advantage of ReRAM-based PIM to alleviate memory bottlenecks while preventing frequent rewrites on ReRAM crossbars.

3. HeTraX achieves up to 5.6× speedup and improves energy-delay-product (EDP) by 14.5× when compared to existing transformer accelerators. Notably, HeTraX exhibits versatility across transformer architectures and shows consistent speedup while being thermally feasible.

## 2 RELATED WORK

Several hardware architectures have been proposed to speedup transformer inference. These architectures include a diverse range of implementations including FPGA, PIM, and architectures using different combinations of non-volatile memory (NVM)-based PIM and SRAM blocks. TransPIM is a DRAM-based PIM accelerator with compute units integrated within High Bandwidth Memory (HBM) banks to accelerate transformer inference [4]. TransPIM uses the host device connected via an interposer for non-matrix computations (normalization, activation functions, etc.) that cannot be performed within the DRAM itself. However, this off-loading of computations adds latency overhead since the system is periodically stalled. HAIMA is another DRAM-based accelerator that adopts a hybrid strategy, incorporating SRAM units for dynamic computations in self-attention and DRAM for multiplications involving large weight matrices [5]. Similarly, H3D-Transformer proposes a hybrid architecture consisting of FeFET, SRAM, and TPU cores stacked vertically via TSV in a 16-tier system [6]. However, it is important to note that HAIMA, TransPIM and H3D-Transformer demonstrate performance gain without taking into consideration the power density and the associated thermal challenges of computing in memory specifically in a 3D architecture. In contrast, HeTraX adopts a holistic view to undertake joint performance-thermal-accuracy tradeoff study.

Alternatively, ReRAM-based PIM accelerators have been proposed to accelerate transformer inference [7]. However, as mentioned earlier, ReRAM-based PIM as a standalone solution is not suitable for transformer inference given the write endurance problem when used for self-attention computation. Xformer tries to address the ReRAM write problem by introducing a hybrid accelerator that uses ReRAM-SRAM arrays [8]. Xformer maps more frequently updated computation kernels to SRAM arrays. However, Xformer is specific to self-attention only and does not consider the end-to-end transformer architecture. Some recent works also propose approximation techniques such as pruning, and sparsification for accelerating transformer inference [1]. However, these investigations are agnostic to specific hardware platforms. To summarize, existing accelerators fail to provide an end-to-end system for transformer acceleration or suffer from notable limitations in terms of system latency or thermal bottlenecks arising due to high power density. In this paper, we fill this gap in the state-of-art by proposing a heterogeneous 3D architecture to accelerate end-to-end transformer models.

## 3 TRANSFORMER MODEL

In this section, we explain the computation kernels of transformers and related architectural variations. The transformer architecture is composed of multiple sequential layers of encoder/decoder blocks, as shown in Fig.1. Each of these blocks consists of two major functional modules: MHA and FF network [1].

**Transformer Computation Kernel:** The input text data to the transformer model is initially tokenized, and subsequently mapped to vectors within an embedding matrix to generate the word embeddings ($I_{emb}$). The word embedding along with positional encodings of the input sequence serve as transformer-readable inputs ($X$) to the MHA module in the model. The MHA computation comprises four computing kernels, outlined in Table 1 (*MHA-1,2,3,4*). The positional encoded embeddings are first multiplied by learned weight matrices (distinct for $h$ heads) to obtain the Query $Q_i$, Key $K_i$, and Value $V_i$ matrices (*MHA-1*). Afterward, the dot products of $Q_i$ and $K_i$ vectors are scaled and transformed into attention scores ($S_i$) using a softmax function (*MHA-2*). The weighted sum of the $V_i$ vectors based on the attention scores produces the context-aware representation ($O_i$) for each word (*MHA-3*). Finally, the outputs from different heads are concatenated and linearly transformed with the weight matrix $W^O$ to create the output ($H_m$) for MHA (*MHA-4*). A layer normalization is applied following MHA (*L-1*), to generate the input ($M$) for the FF. The FF network consists of two computation kernels, as shown in Table 1. The context-aware representation computed through MHA is transformed using a learned weight matrix ($W^{F1}$), followed by an element-wise activation function to generate ($X^1$) output (*FF-1*). This computation is repeated with another transformation using the ($W^{F2}$) weight matrix (*FF-2*). Finally, the output of the FF network ($X^2$) is layer-normalized to ensure stability. The resulting output serves as the input embeddings for the subsequent blocks in the model.

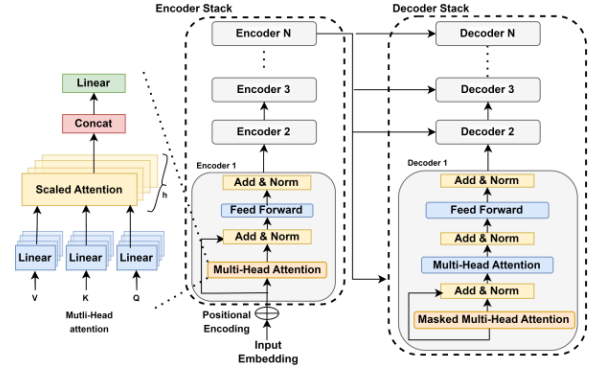

**Fig. 1: Transformer model architecture consisting of encoder and decoder stack of length $N$.**

The original transformer model with encoder-decoder blocks is designed for machine translation tasks [1]. However, the transformer architecture is continuously evolving with structural variations tailored to other NLP tasks. Notable architectural variations include transformers that are exclusively composed of decoder or encoder blocks. This effectively divides the model in half, leading to reduced computational requirements. Additionally, numerous architectural refinements have also emerged within the encoder/decoder block. One such advancement is the introduction of Multi-Query Attention (MQA). MQA, unlike the standard MHA, utilizes the same K and V values across heads while assigning distinct Q vectors. MQA is specifically designed to enhance the computational efficiency of attention mechanisms. Another variation is the parallel attention framework, which involves training the model with both the attention and FF layers operating concurrently. We consider all the different variations (encoder/decoder-only, MQA, and parallel attention) to propose a model-agnostic accelerator for transformers.

**Table 1: Transformer Computational Kernels**

| INPUT | $X = I_{emb} + Positional\ Encoding(I_{emb})$ |
|---|---|
| **Multi-Head Attention (MHA)** | |
| MHA-1 | $Q_i, K_i, V_i = XW_i^Q, XW_i^K, XW_i^V$ |
| MHA-2 | $S_i = Softmax(QK^T)$ |
| MHA-3 | $O_i = V_i S_i$ |
| MHA-4 | $H_m = concat(O_i)W^O$ |
| L-1 | $M = LayerNorm(X + H_m)$ |
| **Feed-Forward (FF)** | |
| FF-1 | $X^1 = GeLU(MW^{F1})$ |
| FF-2 | $X^2 = GeLU(X^1 W^{F2})$ |

## 4 HeTraX ARCHITECTURE

In this section, we introduce the proposed HeTraX architecture and present the design optimization framework.

### 4.1 3-D Heterogenous Architecture

The HeTraX architecture consists of vertically integrated tiers, each with either SM-MC or ReRAM cores. Fig. 2 shows the proposed architecture. Through Silicon Via (TSV)-enabled vertical links are employed to connect the planar tiers. The ReRAM cores consist of crossbars and necessary peripheral circuits (ADC, DAC, eDRAM) to perform matrix multiplication. SMs within the system have tensor cores to enable mixed-precision computing. MCs control the data flow to and from SMs and incorporate a last-level cache. Additionally, MCs facilitate data exchange with DRAM. SM-MC tiers perform the MHA computation since it involves computation with dynamically changing $K, Q, V$ values. The resulting activations serve as the input for the FF network. We execute the FF on the ReRAM tier, given the matrix multiplication with learned weights $W^{F1}$ and $W^{F2}$ (stationary weights). Afterwards, the output of FF network becomes the input for the subsequent block. The 3D system helps facilitate the efficient data exchange between the heterogenous cores and minimize the communication latency.

### 4.2 Performance Optimization

The overall achievable performance of the manycore architecture depends on the computational kernel to core mapping and the associated network-on-chip (NoC). We first maximize the core utilization and determine the optimal core mapping. Subsequently, we design the NoC for HeTraX.

The transformer inference necessitates a substantial memory footprint. Hence, it is imperative to ensure that the architecture under consideration is not underutilized and memory-bound. Nearly two-thirds of the matrix multiplication operations involved in transformer inference are attributed to the FF network [1]. This is predominantly due to the size of the hidden layer, with dimensions 4x larger than the dimensions of all attention heads combined [1]. The adoption of ReRAM-based PIM for the FF computation enables us to prevent the costly repeated off-chip memory accesses that would have been necessary if computation were carried out on the SM cores. However, the FF weights stored on ReRAM crossbars are updated once per layer. Consequently, loading weights from DRAM and updating ReRAM crossbars can take significant time and result in stalls, given the high write latency of ReRAM cells. Therefore, the weight values are updated during the execution of MHA, thereby hiding the write latency. Similarly, MC loads the weights for the MHA during the FF computation. An industry standard interface protocol (DFI) is utilized between the DRAM and MCs [9]. It generates all handshake signals with precise timing requirements necessary for interfacing.

**MHA:** We employ the classical technique of tiling for matrix multiplication in SMs, where blocks of input data are loaded from DRAM to MC. Each SM generates $K, Q, V$ for a specific head with the weight values accessed through MC. Further, we adopt the method of fused score and softmax calculations. The SMs compute score for input sequence blocks sequentially, and the output matrix is generated row-wise over the sequence length. Afterward, the softmax values are computed online for the blocks of rows and appropriately scaled to get the actual softmax for the attention head. A key advantage of this approach is that the attention values are computed without the need to write intermediate matrices back to DRAM, preventing frequent DRAM reads and writes.

**FF:** We map both the weight matrices of the FF network ($W^{F1}$ and $W^{F2}$) to the ReRAM tier with the weights spatially partitioned across the ReRAM cores. This ensures pipelined matrix multiplication with activations flowing unidirectionally from the neural layer $L_i$ to neural layer $L_{i+1}$.

**NoC:** We need to maximize the NoC throughput to maximize the overall performance gain. Given the unidirectional dataflow within the ReRAM tier, core placement as well as associated inter-core links can be determined offline to prevent long-range traffic exchange [2]. Therefore, the inter-core links on the ReRAM tier are not part of the optimization. However, ReRAM cores exchange data with DRAM and the SM-MC tier through vertical links. Hence, we need to consider the traffic through inter-tier vertical links during optimization for the fixed ReRAM core placement. Further, each SM operates in parallel during the MHA computation with data accessed through MCs. The number of SMs typically exceeds that of MCs, thereby giving rise to many-to-few and few-to-many traffic patterns. Additionally, the output from different heads is concatenated which results in a many-to-one traffic pattern. A mesh-based NoC is not well suited for the traffic generated within HeTraX [10]. Hence, following existing work, we design a suitable NoC for the many-to-few traffic generated [10].

Let each candidate design ($\lambda$) corresponds to a specific configuration (placement of cores and links) in the design space. The NoC throughput can be maximized by minimizing the mean $\mu(\lambda)$ and standard deviation $\sigma(\lambda)$ of link utilization [10].

$$\mu(\lambda) = \frac{1}{L} * \sum_{k=1}^{L} u_k \, , \ \ \sigma(\lambda) = \sqrt{\frac{1}{L}\sum_{k=1}^{L} (u_k - \mu(\lambda))^2} \quad (1)$$

Here, $u_k$ is the expected link utilization of link $k$ and $L$ is the total number of links.

### 4.3 Thermal Optimization

3D architectures have high power density and are prone to thermal hotspots. Consequently, it becomes necessary to consider the peak temperature of the system when optimizing for performance. The peak temperature of each core can be estimated using the approximate thermal model, which determines the temperature using horizontal and vertical heat flow [11]. The vertical heat flow can be determined by dividing the system into vertical columns. The temperature of a core located at layer $k$ from the sink due to the vertical heat flow is given by:

$$T(n,k) = \sum_{i=1}^{k}\left(P_{n,i}\sum_{j=1}^{i} R_j\right) + R_b \sum_{i=1}^{k} P_{n,i} \quad (2)$$

where $P_{n,i}(t)$ is the power consumption of the core at layer $i$ from the sink in the vertical column $n$, $R_j$ is the thermal resistance in the vertical direction, and $R_b$ is the thermal resistance of the base layer [11]. The horizontal heat flow is represented through the maximum temperature difference in a layer.

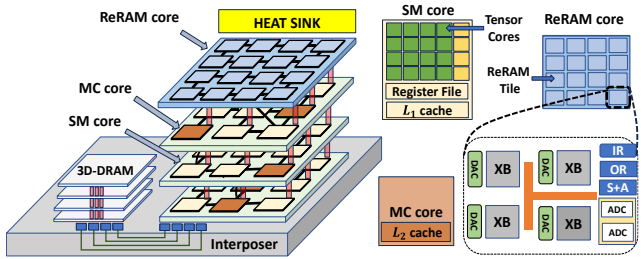

**Fig. 2:** HeTraX architecture consisting of 3 SM-MC tiers and 1 ReRAM tier. This figure is for illustration purposes only.

$$\Delta T(k) = \max_{n} T_{n,k} - \min_{n} T_{n,k} \quad (3)$$

Combining both horizontal and vertical heat flow models, the objective becomes minimizing the worst-case temperature.

$$T(\lambda) = \left(\max_{n,k}\{T(n,k)\}\right)\left(\max_{k}\{\Delta T(k)\}\right) \quad (4)$$

Further, it should be noted that ReRAM cores have different thermal sensitivity when compared to the CMOS-based SM-MC cores. ReRAM cells store weights using conductance, which varies with temperature [3]. Any thermal fluctuations within the ReRAM cells can introduce noise, potentially resulting in erroneous computations and compromising the inference accuracy of the underlying model [3]. The effect of thermal noise on the ReRAM conductance value can be represented by the following.

$$Noise(\lambda) = N\left(0, \frac{\sqrt{4G.K_b.T_{ReRAM}.F}}{V}\right) \quad (5)$$

Here $G$ represents ReRAM ideal conductance, $K_b$ is the Boltzmann constant, $F$ denotes the operating frequency, and $V$ is the voltage difference across the two ReRAM terminals [3]. We consider the temperature of ReRAM cores in the context of the ReRAM noise as the optimization objective.

### 4.4 Overall Design Optimization

Considering the performance, thermal, and ReRAM noise objectives, we aim to find a suitable placement of cores, routers, and links while reducing the overall system temperature and suffering no model accuracy loss due to ReRAM noise. The design of the proposed heterogeneous architecture given the objectives can be formulated as a multi-objective optimization (MOO) problem. We represent the MOO formulations as follows:

$$\lambda^* = MOO(objectives = \mu(\lambda), \sigma(\lambda), T(\lambda), Noise(\lambda)) \quad (6)$$

where $\lambda^*$ is the Pareto optimal set of designs. Given the power constraints, the maximum number of links as well as the number of ports per router can at most be equivalent to a 3D mesh during design space exploration. In this work, we use an ML-guided MOO algorithm, MOO-STAGE [10]. MOO-STAGE has been shown to outperform more conventional MOO algorithms based on simulated-annealing (AMOSA) and other heuristics search especially for a high number of design objectives as is the case shown in $Eq.$ (6). We solve the MOO problem to obtain the Pareto optimal designs with suitable core, link, and router placements. Finally, we perform cycle-accurate simulations to evaluate the Pareto optimal set to determine the best design.

## 5 EXPERIMENTAL RESULTS

In this section, we present an experimental analysis of the proposed HeTraX architecture. First, we describe the experimental setup used for evaluation. Next, we assess the optimization undertaken for HeTraX. Finally, we present a detailed performance-thermal evaluation and EDP analysis with respect to existing accelerators.

### 5.1 Experimental Setup

We consider five transformer models in our experimental evaluation: BERT-Tiny, BERT-Base, BERT-Large, as well as BART-Base, and BART-Large [1]. Furthermore, we also consider different types of transformer architectures, including Encoder-Decoder, Encoder/Decoder only, MQA, and parallel MHA-FF models [1]. All models use 16-bit precision for computation. The 3D system under consideration is structured with four planar tiers, each with size 10 mm x 10mm. We have considered 21 SMs and 6 MCs distributed across three tiers, while 16 ReRAM cores are placed in one tier. Within each SM-MC tier, 9 cores are placed in a 3 x 3 grid pattern, and the ReRAM tier has 16 cores in a 4 x 4 grid. Fig. 2 shows an illustration of the HeTraX architecture. The specifications of the ReRAM, SM, and MC cores used are shown in Table 2. Note that this is an example system size for performance evaluation. The proposed design methodology is applicable to any other configurations. We obtain the application traces using *nvidia-smi* with Nvidia Tesla V100 GPU to evaluate the performance. We use *AccelWattch* to obtain the power profiles of the SM-MC core [12]. For ReRAM cores, we employ a modified *NeuroSim* to obtain the area, latency of all on-chip buffers, and peripheral circuits [13]. We use the cycle-accurate *BookSim2* simulator for implementing the NoC to connect the cores [14]. The inputs to the *BookSim2* are the connectivity between routers and the inter-core traffic traces for the transformer models. We employ a standard NoC flow control mechanism (FIFO-based) for synchronization. Finally, we use the HotSpot tool to conduct steady-state thermal analysis and obtain thermal conductivity values for TSV integration from [15].

We compare the performance of the HeTraX with two recently proposed state-of-the-art transformer accelerators, HAIMA and TransPIM [4] [5]. In our performance evaluation, we assume that the model parameters are available in DRAM before inferencing, and we account for the timing overhead associated with loading weights from DRAM to the MC. Additionally, we do not consider the ReRAM-only architecture, ReTransformer in this comparative performance evaluation due to the endurance concern when used for MHA computation [7]. To illustrate, consider the BERT-Large model processing a single input sequence of length $n = 1024$, where each attention head is mapped to a unique ReRAM core. We need $\sim 5*10^4$ rewrite operations to ReRAM cells. Notably, the number of necessary rewrites increases with the sequence length due to dynamic matrix multiplications. Consequently, the MHA computation on ReRAMs can quickly approach the ReRAM endurance limit ($10^6 - 10^9$ [3]). This issue becomes worse given the trend of continuous increase in sequence length. Contrarily, the FF network computation on ReRAM is independent of sequence length. Due to the fixed and limited number of updates, we employ ReRAM to exclusively implement the FF network in HeTraX.

**Table 2: HeTraX Architecture Specifications**

| ReRAM Tier: 16-cores, 16-tiles per core | |
|---|---|
| ReRAM Tile | 96-ADCs (8-bits), 12×128×8DACs (1-bit), 96 crossbars, 128×128 crossbar size, 2-bit/cell resolution, 10 MHz, 0.34 W, 0.37 mm$^2$, Tech node – 32 nm [2] |
| **SM-MC Tier: 9-cores** | |
| SM core | Volta architecture, 8 Tensor core, 64 KB register file, 96 KB L$_1$ cache, 1530 MHz clock frequency, 9.1 mm$^2$, Tech node – 12 nm [12] |
| MC core | L$_2$ cache – 512 KB, 3.2 mm$^2$, Tech node – 12 nm [12] |
| **TSV Parameters** | |
| Diameter – 5 μm, Via Height – 25 μm, Cap. – 37 fF, Res.- 20 mΩ [17] | |

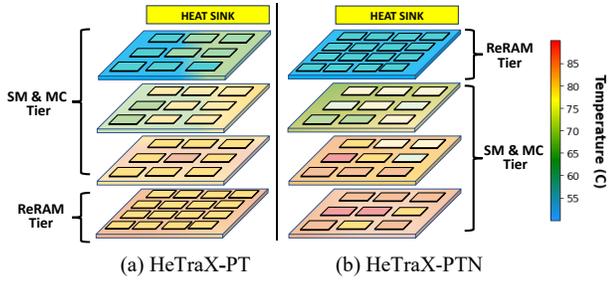

(a) HeTraX-PT    (b) HeTraX-PTN

Figure 3: Comparison of physical placement of cores in HeTraX with/without noise as an optimization objective.

## 5.2 HeTraX Optimization

This section presents an evaluation of the HeTraX optimization framework. First, we validate the necessity of incorporating ReRAM noise as an optimization objective. We conduct a thermal analysis of HeTraX, wherein optimization is pursued under different objectives. The MOO-STAGE algorithm is run for 50 epochs with 10 perturbations from the same starting point. Fig. 3(a) illustrates the core placement when we undertake only joint performance-thermal (PT) optimization similar to existing work on 3D manycore architectures. The core placement under PT optimization places the ReRAM core at the bottom (farthest from the heat sink), given that the SM-MC tier dissipates more power as compared to the ReRAM tier. The overall peak system temperature is 78°c. Conversely, Fig.3(b) depicts the core placement considering joint performance-thermal-ReRAM noise (PTN) objectives. The core placement changes significantly with the addition of noise as an optimization objective in PTN. In the PTN scenario, the placement of the ReRAM tier is nearest to the heat sink. This shift in placement is due to the sensitivity of ReRAM to thermal noise as compared to CMOS-based SM-MC cores and the addition of noise as an optimization objective. The peak system temperature is slightly higher in the PTN scenario (81°C) due to the placement of SM cores at the bottom tier away from the heat sink.

Now, we investigate the impact of ReRAM noise on the model accuracy under the two discussed optimization scenarios. We quantify the effect of noise through $Eq.$ (5), utilizing ReRAM tier temperature. We evaluate the model using widely used GLUE benchmarking tasks. The GLUE benchmark is a collection of 9 tasks specifically designed to evaluate and compare the performance of transformer models [16]. We show the accuracy results considering two NLP tasks, SST-2 and QNLI from the GLUE benchmark for brevity. Fig. 4 presents the accuracy results on both the tasks under three scenarios: HeTraX-Ideal (baseline with no thermal noise), HeTraX-PT, and HeTraX-PTN. We observe that HeTraX-PTN does not suffer from any accuracy loss.

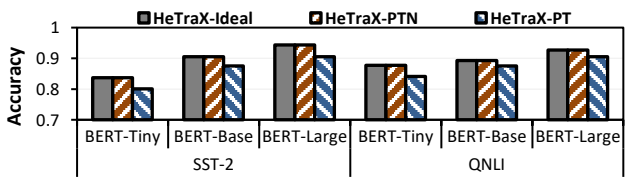

Figure 4: Model inference accuracy with/without the adoption of ReRAM noise as an optimization objective.

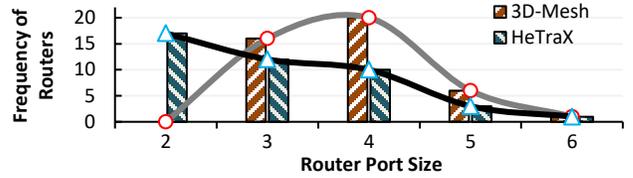

Figure 5: Comparative evaluation between 3D-MESH NoC and HeTraX NoC assessing router port configuration.

This is attributed to the lower temperature of the ReRAM-tier (57°C), where thermal noise remains confined within the quantization boundaries of the ReRAM cells. Conversely, HeTraX-PT observes up to 3.3% accuracy loss. This reduction in accuracy is linked to the elevated temperature at the ReRAM tier (78°C), despite the peak system temperature being lower than that of HeTraX-PTN. These findings underscore the importance of considering ReRAM noise as an additional optimization objective.

Now, we evaluate the NoC of HeTraX against a 3D-MESH NoC design considering optimized core placement obtained under PTN MOO optimization. The router port size directly relates to the NoC performance with bigger routers implying more links. Fig. 5 presents the number of routers with a specific number of ports for the two NoC topologies. We observe a lateral shift to lower router port count in HeTraX as compared to mesh NoC design. This shift suggests that the NoC in HeTraX has relatively smaller routers and a reduced number of links, which directly influences the NoC performance and energy consumption. Overall, these results validate the effectiveness of HeTraX's joint performance-thermal-accuracy optimization. The optimization determines suitable core and link placement to achieve high performance while being thermally feasible and suffering no accuracy loss.

## 5.3 HeTraX Performance Analysis

Now, we compare the end-to-end latency of HeTraX with the baselines TransPIM and HAIMA. Fig. 6(a) shows the normalized execution time for each computational kernel of the encoder-only BERT-Large model as an example. We observe that HeTraX outperforms both HAIMA and TransPIM and achieves speedup for each computational kernel within the transformer model. The SMs efficiently accelerate MHA computation, and the ReRAM layer computes the FF layer in a highly parallel manner. Additionally, HeTraX benefits from the fused score and online softmax calculation on SM cores. In contrast, HAIMA and TransPIM rely on an additional host for softmax, which prevents online execution and results in repeated data exchange with the host. However, as highlighted in Section 2, HAIMA and TransPIM did not consider thermal effects while designing the architecture. For instance, HAIMA's configuration suggests integration of up to eight compute units per bank, with each compute unit dissipating 3.138W of power. Under the assumption that all compute units in a bank are operating concurrently as suggested in HAIMA, the power density of the HBM bank will be around 8W/mm² (*16x higher than modern GPUs*) given the standard HBM2 die area of 53.15 mm² for 16 banks. Similarly, TransPIM consists of 8 stacks of HBMs connected through TSV. The thermal resistance increases as we move up in the stack and away from the heat sink. Consequently, additional power dissipation through compute units in banks can increase DRAM temperature quickly and create hotspots.

**Performance-Thermal Evaluation**: Fig. 6(b) presents the overall performance as well as steady state system temperature with the same model size as BERT-Large along with different transformer architectures to conduct a thorough performance and thermal analysis. Notably, we observe that both HAIMA and TransPIM exhibit elevated temperatures, with a minimum temp of 120°C for any transformer architecture. The maximum temperature reaches 142°C in the case of the fused MHA-FF model where MHA and FF layers are computed parallelly. Note that the maximum temperature threshold for DRAM is 95°C beyond which it results in data loss or corruption. These results underscore the thermal infeasibility of both HAIMA and TransPIM, necessitating the exploration of dynamic voltage frequency scaling techniques to make them practically viable. However, the exploration of such techniques falls beyond the scope of the current work. Conversely, HeTraX proves to be thermally realizable, consistently achieving speedup across various transformer architectures. The speedup is similar for Decoder-only and Encoder-Decoder models. MQA achieves slightly more speedup due to its reduced memory bandwidth requirement [1]. The speedup is maximum for parallel attention where vertical integration of heterogenous tiers helps with efficient data exchange during parallel MHA and FF computation. In summary, HeTraX demonstrates superior performance and temperature characteristics when compared to the baselines.

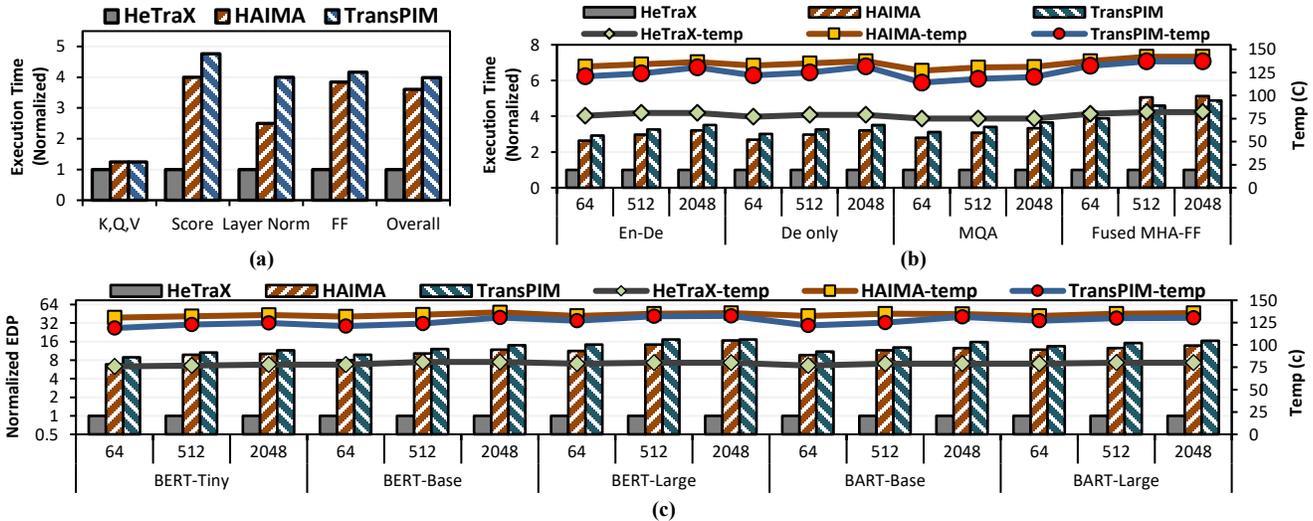

Figure 6: Normalized execution time and temperature when compared to the baseline HeTraX, (a) each computation kernel within BERT-Large, (b) different transformer architectures maintaining uniform model dimensions, and (c) different transformer models with varying sequence length.

**EDP Evaluation**: Fig. 6(c) illustrates the normalized EDP along with steady state temperature for HAIMA and TransPIM with respect to HeTraX considering varying real-world transformer models and input sequence lengths. We consider EDP as the relevant metric as it captures both performance and energy in a single term. It is evident that HeTraX outperforms the baselines exhibiting increased EDP gains as the transformer model size and input sequence length increases, thereby showing scalability. For instance, the EDP of HeTraX is an order of magnitude better (14.5×) than HAIMA for the BERT-Large model with $n = 2056$.

## 6 CONCLUSION

In this paper, we consider the complex interactions among various types of computing elements to design a 3D manycore architecture called HeTraX for accelerating end-to-end transformer models. HeTraX uses SM-MC cores for MHA and ReRAM cores for the FF network and integrates them vertically for faster data exchange. Experimental results demonstrate that HeTraX outperforms the existing state-of-the-art in terms of execution time and EDP while ensuring thermal feasibility.

## ACKNOWLEDGMENTS
This work was supported, in part by the National Science Foundation (NSF) under grants CNS-2308530 and CNS-1955353.